\documentclass[12pt]{article}
\usepackage{mathtext}
\usepackage{amsfonts}
\usepackage{amssymb}
\usepackage{amsmath}
\newcommand{\G}{\Gamma}

\renewcommand{\d}{\delta}

\newcommand{\g}{\gamma}

\newcommand{\D}{\Delta}

\newcommand{\e}{\epsilon}

\newcommand{\ar}{\longrightarrow}

\newcommand{\s}{\sigma}
\newcommand{\la}{\lambda}
\renewcommand{\a}{\alpha}
\begin{document}
\title{Simulation of quantum dynamics via classical collective behavior}
\author{Yu.I.Ozhigov\thanks{The work is supported by the Fond of NIX Computer Company (grant \# F793/8-05), and INTAS (grant 04-77-7289). e-mail: 
ozhigov@cs.msu.su} \\[7mm]
Moscow State University,\\
Institute of physics and technology of RAS 
} 
\maketitle
\begin{abstract}
The simple algorithm for the simulation and visualization of non relativistic quantum dynamics is proposed that is based on a collective behavior of classical particles. Any quantum particle is represented as the swarm of its classical samples which interact by simple rules including emission and absorption of samples of tied photons. The quantum dynamics results from the collective behavior of such a swarm where the eigenstates are treated as the equilibrium states relatively to emission-absorption of photons. The entanglement is treated as a correlation between samples of the different swarms that is stored in the space-time part of the model inaccessible for a user. The amplitude is always grained. The Coulomb field between quantum particles is simulated, analogously to free flow of quantum package, by the point wise interaction between its samples and scalar photon samples which propagate by diffusion. It gives square root speedup in comparison to each-with-each method. This method obviously includes decoherence and admits the natural generalization on the QED of many particles with the linear computational cost. 
\end{abstract}
\newpage
\section{Introduction}

Many particle problems represent the natural area for the applications of quantum theory, and just here it meets the unexpected and principal difficulty. At first glance this difficulty seems not fatal because it is connected with the complexity of the description of many body problems on classical computers, e.g., with the applications of quantum physics. But this viewpoint is wrong. In the more detailed consideration the more and more depths of this difficulty become apparent. We can now assert with certainty that this difficulty is so serious that it factually establishes the limitation of the applicability of quantum theory itself, so that the adequate description of many body systems requires the substantial revision of its basement: namely, of its mathematical apparatus. The difficulty is that the dimensionality of the space of states of many body systems grows as the exponential when the number of its particles increases. The natural solution would be the creation of a quantum computer for the simulation of many body quantum systems. 
\footnote{The idea of quantum computer was in the air and has been expressed almost simultaneously by few people: Benioff, Manin, later - Deutsch, but Feynman was the first who formulated it clearly as applied to the simulation of quantum many body problems, he also reformulated quantum theory on the intuitive language of paths (that corresponds for our aim better than the traditional Shroedinger form). Feynman, of course, did not state the question about the limitation of applicability of Hilbert spaces apparatus for that it was not necessary experimental data at that time.}. 
The idea of the creation of quantum computer obviously belongs to the great ideas of physics. The valuable part of physicists (including the author of this paper) is somehow involved in this work. The quantum computer is the first principle prediction of quantum theory that follows immediately from the mathematical formalism of tensor products of Hilbert spaces. Hence, the creation of such a device must be treated not as the technological task but as the decisive experiment on the checking of quantum theory in the many particle case. It worth noting that quantum theory has been never verified in many particle case; all the results experimentally confirmed at the moment are obtained by the different reduction to the one particle case
\footnote{The single exclusion is the attempts of the direct computation of the energy of bipartite systems via wave function $\psi (r_1,r_2)$, for example, for helium atom (see below). The attempts to solve many body problems analytically always ended by ambiguous results: for example, the systematic consideration of evolution of two photons and one electron in QED leads to the divergence of rows for the amplitude.}
The factual implementation of quantum theory in the many body area, e.g., the practical, computational consideration of the wave functions of the form $\psi (r_1,r_2,\ldots,r_n)$ for large $n$ cannot be realized neither analytically nor numerically due to the phenomenon of quantum speedup of classical computations that was discovered in ninetieths. The sense of this phenomenon is as follows. Accordingly to Hilbert formalism, the states of system with $n$ particles belong to the tensor product ${\cal H}_1\bigotimes{\cal H}_2\bigotimes\ldots\bigotimes{\cal H}_n$ of spaces ${\cal H}_j$ for one particle states (for simplicity we assume that all particles are distinguishable; otherwise we should consider Fock space of occupation numbers with bosons or fermions symmetry instead of Hilbert space that complicates the representations without changing the essence). 
And conversely, any normalized vector in this space represents some physically realizable state of $n$ particle quantum system. 
The dimensionality of this space equals the product of dimensionalities of all ${\cal H}_j$, e.g., it grows rapidly with the growth of $n$. It means that there exist such states that amplitudes of some basic states in it are very small. It is impossible to check the presence of such small amplitudes by the straightforward measurements because the corresponding probabilities are negligible. 

But there exist the special forms of  many body system evolutions when these small amplitudes constructively add in the huge quantities and give the observable value in so short time that no classical method of computations (in particular, no analytical calculations) can keep pace with this process. 
There are the quantum computations. It is important that there exists the exact criterion separating a quantum computer from classical ones: that is the speed of solution of some computational tasks (for example, the search problem). 
We thus have the project of decisive experiment checking the truth of quantum theory in the many body area. In the present time all the attempts to build at least a few qubit reliable quantum computers (let along a scalable one) failed. Of course, one cannot yet conclude that the quantum theory cannot be applied to the many particle tasks. But it makes the building of alternative formalism for many body problems actual. Such formalism must embrace all known quantum physics including entangled states but a scalable quantum computer. The single possible mathematical formalism for such extrapolation of quantum, theory to the many particle case is the classical algorithm theory. The simplest idea is to introduce the grain or the quantum of amplitude (the minimal nonzero module of amplitude) for quantum states and to consider the corresponding reduction of quantum theory, which was proposed in the work \cite{Oz}\footnote{It makes our model dependent on the choice of basis. This cost seems not too large because algebraic universality is a feature of Hilbert spaces formalism and it does not follow straightforwardly from the experiments.}. It turns that this simple trick gives the unified description of quantum evolutions without division to the unitary evolutions and the measurements (including decoherence). The Born rule for quantum probability follows from the hypothesis of amplitude quantum immediately. But this method is not yet ready for practical implementations. It is based on the notion of many-body wave function and for the computer realization it requires the immediate using of the amplitude quantum which real value can turn too small for this. It causes over expenditure of the computational resources and thus allows solving mainly one particle problems.  

In this work we consider the version of algorithmic approach (see. \cite{Oz}), based on the collective behavior of classical many body systems. This version can be named swarm method, because it uses the representation of a quantum particle as a swarm of its samples, each of which is a classical particle and can interact with others by simple rules. Any set of identical quantum particles is represented by the corresponding joint swarm. This method admits the easy passage from the classical description of particle to the quantum and vice versa, it is scalable and convenient for the creating of computer programs. In this version only EPR type of entanglement can be adequately represented that is the essential reduction of Hilbert quantum formalism. In the framework of swarm method the fundamental interaction between charged particles and photons can be interpreted, that means that this method can embrace QED, and probably some other types of interactions. 

\section{General features of algorithmic approach: user and administrative parts of model}

We make some overview of algorithmic approach. The main tool here is classical algorithms, not the analytical technique. The final result is represented by the algorithm simulating the dynamics of system at hand, not by the value of some parameters. From the theory of algorithms it is known that the work of algorithm cannot be predicted beforehand. The only way to learn how it works is to launch it and wait when it returns the result. The single universal form of the evaluation of results is thus the observation of video film which is the visualization of the work of the simulating algorithm. The usual physical methods like finding Hamiltonians, eigenvalues and eigenvectors etc., will obtain the status of auxiliary instruments, checkpoints or heuristics in building of the final dynamical picture of evolution. This results in the division of the model to the user and the administrative parts. The first one is accessible for user (it inserts the initial data and shows the video film). The second part is inaccessible for a user; it prepares the film through the simulating algorithm. 

The physical sense can be acquired for objects containing in the user part only, which can be observe immediately. As for the administrative part, its details have no physical sense. The administrative part is needed for the film preparation only and it cannot be associated with some object in the real world. In particular, the time spent to the film preparation is not the physical time\footnote{Some authors use the term “hidden time” of quantum theory, see, for example, \cite{KB}.}. The necessity of the administrative part follows from the existence of quantum non locality, e.g., entanglement of spatially separated particles. The simulation of such systems cannot be fulfilled in the real time mode, the film must be prepared beforehand to be able to work at the both parts of a spatially separated system in an entangled state. Here it would be naturally to provide by the “free will” only users, but not the administrative part of the model\footnote{I do not intend to go into the details of the philosophic character.}. For example, if we need a source of random numbers we can include it to the administrative part as some program generating pseudo random numbers. The seeming absolute character of quantum randomness will thus result from that we do not know the parameters of this generator which lies in the administrative part. We also should not be excited by the possibility to “learn” these parameters by some non authorized way and then use it for our technologies. The right simulating algorithm produces the film which is independent from the individual parameters of randomness generator. The radical viewpoint is that the randomness generators are needed for the debugging of the true model embracing the huge number of particles, for example, in biology. We will use the generators of randomness because our immediate aim is the physics of many-body systems where the randomness is the fundamental element of evolution. 

The bipartite structure of the model would make no sense without the imposing of strict limitations to the abilities of the administrative part. This limitation is only one – our model must represent the classical algorithm realizing effective (polynomial complexity) algorithms. If we consider the whole Nature as such a model, we would treat ourselves as its users and the administrative part as the internal structure of the world which immediate access is forbidden for us, for example, by some password. Despite of the closed character we can make one conclusion about its abilities: it cannot realize fast quantum algorithms. This feature of the model allows the contra version and it can be experimentally checked. Such experiment is the building of a scalable (10000 or more qubits) quantum computer. 
The creation of such device would mean the crash of algorithmic approach. But no failure in quantum computer building can serve as the proof of correctness of algorithmic approach. The correctness of complex films can be the unique criterion here. 

The main aim in the algorithmic approach is the creation of films reflecting the real evolutions of many-body quantum systems. Just this aim we keep in mind describing our model. As for the correspondence with the standard quantum physics, such a correspondence cannot be complete, because we at first exclude the description of $n$ particle system via wave function of the form $\Psi (r_1,r_2,\ldots ,r_n)$. This description is possible in our model only for the case of the entangled states of EPR type, for example, one photon and one particle, two entangled photons, or many particles which behave as the single particle, etc. In the reality this assumption is not very restricting; we can find substantially exact corresponding with QED that is the subject of future investigations. We here take up only non relativistic approximation. Nevertheless, it will be clear how to find the rules for photon emission-absorption which agree with QED.

\section{Swarm model of quantum dynamics}

Swarm model (or model of collective behavior) consists of that each quantum particle is represented as the set (swarm) of classical particles of the form
\begin{equation}
\label{swarm}
s=\{ s_1,s_2,\ldots,s_k\} ,
\end{equation}
where each particle $s_j$ is called a sample of the considered quantum particle and has the definite space-time coordinates $t(s_j),\bar x(s_j)$ and some parameters needed for the simulation of dynamics. For example, we can introduce for each sample the pointers to the other samples of this quantum particle or the other quantum particles. 
The pointers associated with $s_j$ point to the samples which influence is most essential for $s_j$. In the swarm representation of charged particles space-time coordinates are separated because these particles are classical and their impulses are always connected with the coordinates as $p=m\frac{\D x}{\D t}$. The close coordinates in QED is the good criterion of the essentiality of interaction, and consequently we can manage without the pointers at all. 
The simple swarm representation of $n$ charged particles has thus the form of list $\bar s=(s^1,s^2,\ldots,s^n)$, which components $s^k$ are swarms representing the separate particles. This simple swarm representation is valid for the description of the state of this system in any time instant. In the language of standard Hilbert formalism it means that this state will always has the form $\Psi_1\bigotimes\Psi_2\bigotimes\ldots\bigotimes\Psi_n$, e.g., will be not entangled. This is the fundamental difference between swarm method and Hilbert formalism, because the most essential property of the last is the existence of entangled states. It worth noting that we could simulate the standard formalism via pointers from the samples of charged particles to the other samples of this sort.

If we introduce the operator of free flight for the swarm representation and some simple rules like creation-annihilation of samples, the simple swarm representation is valid for finding of the ground states of a system with $n$ electrons (the anti symmetry of wave function can be reflected in the framework of this representation). But such operator cannot simulate the dynamics of quantum systems even in case of one particle in potential (excluding the case of free particle flow which is simulated by the diffusion of its samples). 

To simulate the unitary evolution induced by Shroedinger equation we need to introduce the samples of the different sort. These samples must 1) move very fast, 2) they must be emitted by the samples of charged particles, and they must transform to the samples of charged particles, and 3) they must store the memory about what sample of charged particle they are emitted from; e.g., there must exist the pointers connecting them with samples of charged particles. 
In view of these properties we call these new samples the samples of tied photons. We can introduce vector photons which direction of propagation is orthogonal to the vector $\bar E$ of electric field, and the scalar photons moving along $\bar E$. At last we can define the conditions of transformation of samples of tied photons to the samples of free photons that is needed for representation of QED.

The pointer of sample $\a$ of particle or photon sample $\g$  is the natural number $p(\a )$ or $p(\g )$, denoting the number of the other sample - of photon or particle correspondingly. The pointers must satisfy the reciprocal condition: $p(\g_k)=m\ iff\ p(\a_m)=k$.

The swarm representation of $n$ particle system in electromagnetic field has the form of list of vectors
\begin{equation}
\label{gen_swarm}
\bar s_{par} =\{ s^1,s^2,\ldots,s^n\},\ \ \bar\g=\{\g_1,\g_2,\ldots,\g_l\} , \bar g=\{g_1,g_2,\ldots,g_d\},
\end{equation}
where $\bar s_{par}$ is the vector consisting of samples of charged particles with pointers “particle-photon” to the samples of photons which form the vectors $\bar\g$, $\bar g$ of the tied and free photon samples. The evolution of this representation (\ref{gen_swarm}) in time is determined by the sequential operators of the following types: creation of samples of tied or free photons by the samples of charged particles (photon emission), creation or annihilation of the samples of charged particles by the photon samples (absorption) - here the sample of tied photon disappears and the sample of free photon remains, and also creation and annihilation of the samples of charged particles and photons. The interaction of QED can be simulated by this type of interactions. 

From the viewpoint of Hilbert formalism a state of system of $n$ quantum particles we deal with in swarm method will always have the form  
\begin{equation}
|\Psi\rangle=|\Psi_1\rangle\bigotimes |\Psi_2\rangle\bigotimes\ldots |\Psi_n\rangle ,
\end{equation}
where $|\Psi_j\rangle\in {\cal H}_{particles}\bigotimes {\cal H}_{j,\ photons}$ is not entangled state of $j$-th particle and photons of the form
\begin{equation}
\Psi_j= \Psi_{j,\ par}\bigotimes\Psi_{j\ phot}
\end{equation}
$\Psi_{j\ phot}$ is the state of some subset $Ph_j$ of all photon samples so that the sets $Ph_j,\ j=1,2,\ldots,s$ form the division of the set $Ph$ of all photon samples. No entanglement thus arises in the passage to the standard language of Hilbert spaces.   

Here the operator of time evolution for such system has the following form:
\begin{equation}
U_1^{\Psi_2,\Psi_3,\ldots,\Psi_n} \bigotimes U_2^{\Psi_1,\Psi_3,\ldots,\Psi_n}\bigotimes\ldots\bigotimes U_n^{\Psi_1,\Psi_2,\ldots,\Psi_{n-1}} ,
\label{e}
\end{equation}
where $U_j^{\Psi_1,\ldots,\Psi_{j-1},\Psi_{j+1},\ldots,\Psi_n}$ is the superposition of a) unitary operator on $j$-th particle which depends on the states of all other particles and b) creation and annihilation of photon samples leading to the change of the sets $Ph_j$. 
We thus have only not entangled states of particles and photons but an evolution of such states cannot be represented as sequence of one particle operators. The evolution induced by the operator (\ref{e}), cannot be expressed in form of quantum gate array because the application of any entangling operator gives the entangled state. The method of collective behavior for many particles can be thus expressed in terms of tensor products, e.g., here we have the radical difference from the many body Hilbert formalism. Operators of the form (\ref{e}) arise in mean field approximation that is applied in the Hartree-Fock method. This method and the diffusion Monte Carlo method give good approximations of the experimental values of energy for many electron systems (atoms and molecules), with the accuracy within few percents. The most efficient methods of many body computations thus can be well approximated by thhe swarm method. 

The entangled states of the form:
\begin{equation}
\sum\limits_{j}(\la_j)^n|\Psi^1_j\rangle\bigotimes |\Psi^2\rangle\bigotimes\ldots\bigotimes |\Psi^n_j\rangle
\label{shm}
\end{equation}
have satisfactory swarm representation, where for all $k=1,2,\ldots ,n\ \ \ $ $\Psi^k_1,\Psi^k_2,\ldots$ 
is orthonormal system of functions in the space ${\cal H}_k$ of states of particle or photon $k$. Such states generalize the known Schmidt decomposition for bipartite states but only in two particle case ($n=2$) any state can be reduced to them. Swarm representation of the state (\ref{shm}) has the form (\ref{gen_swarm}) where $s^k$ is the swarm representing the state 
\begin{equation}
\sum\limits_{j}(\la_j)|\Psi^k_j\rangle ,
\end{equation}
where pointers are arranged so that they connect samples corresponding to the basic one particle functions $|\Psi^k_j\rangle$ with the same lower index $j$. 

Swarm approach thus principally reduces the possible types of entanglement which exist in tensor products of Hilbert spaces.

\section{Swarm representation of many-body wave functions as alternative to Hilbert formalism}

The principal difference between the method of collective behavior and Hilbert formalism is that in the first method the four dimension space-time is the single space of states and its dimensionality does not grow with the number of particles. It results from that the set of quantum particles is represented by the set of the corresponding swarms which exist in the same space-time. The versions of swarm approach are the known diffusion Monte Carlo method and its modifications like Semenihin method for finding of excited states (\cite{Se}). 

Our aim here is the representation of the quantum dynamics of a system consisting of many charged particles interacting through photons accordingly to QED.  

\subsection{Representation of one particle wave function through collective behavior with tied photons}

We give more universal representation of quantum evolution which can be then generalized to the case of many particles. Shroedinger equation is twice more complex than the diffusion transform because the wave function $\Psi(x,t)$ - is complex-valued. It carries twice more information about the state of swarm than the density $\rho$ from the diffusion equation, because besides coordinates of samples we must know their speeds. 
We are going to describe one particle quantum dynamics in terms of swarm method. The simple considerations show that it is impossible if we are limited by only samples of one particle even if we will divide them to the different types; arguments for that see below. Hence, we must introduce the samples of some other auxiliary particle (tied photon), different from the initial. The role of this auxiliary particle is to ensure the quantum dynamics of the initial one. Here the stationary functions $|\phi_j\rangle$ of Hamiltonian correspond to the equilibrium states $s^{stat}_j$ of the swarm $s$, that means the steadiness of necessary storage of the tied photon samples, whereas in the passage from one stationary state to another we need to change this storage by increasing or decreasing it. It makes reasonable to call this new particle tied photon, despite of that such characteristics of photons as polarization will not be yet included to the description of simple quantum dynamics. In this section the association of new particle with electrodynamics’ photon will have no consequences because we will consider not a real particle but its samples only.  

Shroedinger equation for the wave function $\Psi(x,t)$ has the form

\begin{equation}
i\dot\Psi=-\D\Psi+V\Psi .
\label{shr}
\end{equation}
We represent the wave function in the form 
$$
\Psi(x,t)=\Psi^r(x,t)+i\Psi^i(x,t),
$$
where $\Psi^r(x,t),\ \Psi^i(x,t)$ are its real and imaginary parts. The equation (\ref{shr}) can be then written as the system of two equations
\begin{equation}
\begin{array}{lll}
&\dot\Psi^r&=-\D\Psi^i+V\Psi^i,\\
&\dot\Psi^i&=\D\Psi^r-V\Psi^r.
\end{array}
\label{}
\end{equation}

We establish the necessary connection between Shroedinger and diffusion equations. The diffusion equation has the form

\begin{equation}
\rho_1\dot u=div(p\ grad\ u)-qu+F,
\label{diff}
\end{equation}
where $u(x,t)$ is the density of particles (samples), $\rho_1,p,q,F$ parameters depending on $x,t$ and designing the density of environment, diffusion coefficient, coefficient of absorption and intensity of source of samples correspondingly. The positive coefficient of absorption means that the samples are annihilated in this point with the intensity $|q|$. We assume that $F=0,$ and $\rho_1, p$ has unit value, e.g., we will give the reasonable sense only to the coefficient of absorption $q$, which will be proportional to the potential energy in Shroedinger equation. The equation (\ref{diff}) follows from Nernst law (see. (\cite{Vl}) for the finding of the string of samples through the element of surface $dS$:
\begin{equation}
dQ=-p\frac{\partial u}{\partial\bar n}dS.
\label{ner}
\end{equation}

To reduce the equation (\ref{shr}) to some version of the diffusion equation, we consider the swarm of samples of our quantum particle. We divide these samples to two types: real ($r$) and imaginary ($i$), and in each of these type in turn divide to two subtypes: positive (+) and negative (-). We thus obtain the division of all samples to four types which members will be designated as follows: $ \a^{+,r}_j,\ \a^{+,i}_j, \ \a^{-,r}_j,\ \a^{-,i}_j,$ where $j$ denotes the number of a sample. For the description of stationary states only one type of samples would suffice, because stationary states are determined by the density. For the description of dynamics we really need two types of samples: real and imaginary, but it is more convenient to have four types. We divide the configuration space to the small cubes and let $D(x,t)$ denote the cube containing the point $(x,t)$. 
The total number of samples of the swarm $\bar s$ of a same type containing in some cube $D(x,t)$ will be denoted as  $s^{\sigma ,\eta}(x,t)$, where $\sigma\in\{ +,-\},\ \eta\in\{ r,i\}$. We assume that the speeds of samples are distributed uniformly independently of the type. We intend to represent the wave function evolution as the chain of sequential diffusion we must get rid of signs in equations, and for that we will use the defined types of samples where the swarm approximation of wave function is always found by the formula 
\begin{equation}
\Psi(x,t)_{s}=s^{+,r}-s^{-,r}+i(s^{+,i}-s^{-,i}).
\label{sign}
\end{equation}
This equation does not determined uniquely the division to the positive and negative parts, but only within the addition of a constant to these both parts. 
We then have within the normalization the following approximate equations:
\begin{equation}
\begin{array}{lll}
&\Psi^r(x,t)&\approx s^{+,r}(x,t)-s^{-,r}(x,t),\\
&\Psi^i(x,t)&\approx s^{+,i}(x,t)-s^{-,i}(x,t),
\end{array}
\end{equation}

Shroedinger equation then acquires the following form:
\begin{equation}
\begin{array}{lll}
&\dot s^{+,r}(x,t)&=\D s^{-,i}(x,t)+V(x,t)s^{+,i}(x,t),\\
&\dot s^{-,r}(x,t)&=\D s^{+,i}(x,t)+V(x,t) s^{-,i}(x,t),\\
&\dot s^{+,i}(x,t)&=\D s^{+,r}(x,t)+V(x,t) s^{-,r}(x,t),\\
&\dot s^{-,i}(x,t)&=\D s^{-,r}(x,t)+V(x,t) s^{+,r}(x,t).
\end{array}
\label{swa}
\end{equation}

We will enumerate the types $(+,r),(+,i),(-,r),(-,i)$ by natural numbers $1,2,3,4$ respectively, and will apply to them arithmetic operations in $Z/4Z$ everywhere including indices. We then have already no single density $u$, but the vector-column $\bar u$, which components $u_j\ j=1,2,3,4$ are the densities of samples of types $j$. The rule for transformations of types must correspond to the turn around the center of coordinates in the definite direction, e.g., this rule must be cyclic: $1\ar 2\ar 3\ar 4\ar 1$. 
The system (\ref{swa}) will be then equivalent to the following equation 
\begin{equation}
\dot \bar u=\G (\D u-q\bar u),
\label{diff_trans}
\end{equation}
where the matrix $\G$, expressing our law of transformation of types, and the matrix $g$, inverting the signs of types have the form
$$
\G=
\left(
\begin{array}{lllll}
&0&0&0&1\\
&1&0&0&0\\
&0&1&0&0\\
&0&0&1&0
\end{array}
\right) ,\ \ \ 
g=
\left(
\begin{array}{lllll}
&0&0&1&0\\
&0&0&0&1\\
&1&0&0&0\\
&0&1&0&0
\end{array}
\right) .
$$

We now define the collective behavior giving the solution of equation (\ref{diff_trans}). Here it would not be sufficient to manage with only cyclic rule of transformation of types, because in that case the matrix $\G$ would contain equal and nonzero diagonal elements, whereas in the matrix $\G$ all diagonal elements are zero. This is the reason of including tied photons to the collective behavior. 

We now have four samples of particle: 1,2,3,4 with the same dynamical characteristics and also the samples of tied photons. The samples of particle and photons are denoted by $\a^j$ and $\g^j$ correspondingly. Let each sample of charged particle of the type $j=1,2,3,4$ creates with the constant rate the samples of tied photons of the same type. These samples of tied photons move via diffusion and in the time frame $\d t$ they transform to the samples of the same particle of the type $j+1$. The type transformation then goes indirectly through tied photon samples. The main requirement is that the coefficient of diffusion for tied photon samples must be much nore than for the samples of particle: $p_{phot}\gg p_{part}$. We can assume that the samples of particles do not move at all themselves. This allows to suppress the diagonal elements in the matrix of generalized diffusion that gives the needed mathix $\G$.

We then consider two areas $D_1$ and $D_2$ with the common border through which the diffusion goes. Here the photon samples diffusion goes much faster then the diffusion of samples of particle. In the change of density of any type of samples of particle in the tiem frame $\d t$ two samples of this type make deposit: a) newly created as the transformation of type from the photon samples which either came from the neighboring areas, or were emitted in this area, and b) penetrated by the diffusion. By our supposition about intensities of diffusion, the deposit b) is negligible, and we can consider only deposit a). Applying the formula (\ref{ner}), we obtain that the income of samples of the type $j$ in the unit time frame through the border is 
\begin{equation}
dQ=-p_{phot}\frac{\partial u^{j-1}}{\partial\bar n}dS,
\label{ner1}
\end{equation}
where $u^{j-1}$ is the density of samples of parent type. We treat the potential $V$ as the intensity of creation or annihilation of samples of any type. Using the reasoning from the derivation of the diffusion equation we come to the equation (\ref{diff_trans}). 

The introduced method of description of quantum evolution includes the classical dynamics of particles. It would suffice to have the number of photon samples $A(|s^{r,+}-s^{r,-}|^2+|s^{i,+}-s^{i,-}|^2)^{1/2}$, where $A$ is constant which does not depend on the point of configuration space that prevent the useless storing in memory of the mutually canceling parts of the wave function. Let the carrier of wave function concentrate in some area $D_0$. The distribution of the phase as $\phi(x)=С \bar x\bar p$ then lead to the same result as the mechanic movement of the swarm with the speed proportional to $\bar p$ along the vector $\bar p$. Correspondingly, we can modify the rule for the transformation of photon samples to the samples of particle using the mean speed $v_{av}(x,t)$ of samples of particles which emit the samples of photon in the point $x,t$. The former main rule was that the photon samples of type $j$ transform to the samples of particle of type $j+1$ in time instant $t+\d t$, where $\d t$ is constant. The new rule consists in that the photon samples transform to the samples of particle of the types corresponding to the distribution of amplitudes of the form $exp(\d tip^2/m)$, where $\d t$ is an arbitrary positive number in the segment $(0,\d t_0)$, where the impulse $p=v_{av}m$ is computed basing on the mean speed of samples of particle which emit photon samples. It means that for the fixed $\d t$ in this segment the fraction of the type $j+k, \ k=0,1,2,3$ equals to the corresponding amplitude in the exponential $exp(\d tip^2/m)$. 

The new rule has no influence to the dynamics of swarm if the initial distribution of amplitudes has the form $exp(ipx)$ and $V=0$. In this case the movement is the uniform and has the constant impulse $p$. In the other cases  the influence of this rule is determined by the value $\d t_0$: the less it is, the less the influence will be. 
If we fix the value $\d t_0$, we can find the influence of potential $V$. Let $\D \Psi (\bar\D x)$ be the divergence of wave function values found by the old and new rules for the transformation of photon samples to the samples of particle in the point $x+\bar\D x$. $\D \Psi (\bar\D x)$ will be then proportional to $\sin (\bar{grad}\ V,\bar\D x)$. The divergence is then maximal if the samples of photon move orthogonally to $\bar{grad}\ V$.
It substantiates the following agreement. We assume that a sample of tied photon transform to a sample of free photon if it propagates to the distance exceeding some limit $\D x_0$ from the point of emission towards $\bar\D x$, so that $|\sin (\bar{grad}V,\D x)|>\e_0$, e.g., in the direction close to the normal to the gradient of potential in which the considered particle moves. This agreement is not accurate because we have not defined the constants $\e_0,\ \D x_0$ in the definition, but it can be reformulated in terms of QED if we introduce the vector of polarization of photon which is orthogonal to the photon impulse. Our agreement then will express in swarm terms the rule for finding of amplitude of fundamental process of a photon emission by charged particle (see (\cite{Fe})).

\subsection{Representation of Coulomb field}

The collective behavior defined below has one drawback: it does not explain completely the mechanism of photon exchange and thus it is not appropriate for the extrapolation to QED. Namely, the potential $V(x,t)$ is used without explanation of its generation. It is sufficient for the creation of video film about many different particles interacting by Coulomb (or any other) law, but it cannot serve as a tool for description of emission – absorption of the real photons. But the main deficiency is that the described method cannot be scalable due its huge computational complexity: for $N$ samples to one particle and $k$ particles one step of simulation requires of the order $k^2N^2$ operations because with the assumption of instantaneous propagation of the field the interactions in all pairs of samples for the different particles must be considered. We show how to reduce this complexity to $kN$. For this we assume that the field propagates with the finite speed. In addition, our construction is true for the Coulomb field only. We introduce the mechanism of creation of Coulomb potential $V(x,t)=\frac{\g}{|x-x_0|}$, by the center located in the point $x_0$. 

Let the center emits samples $g_1,g_2,\ldots$ of new type – free scalar photons, each of which is supplied by the sign of charge which has emitted it $sign(g_j)$, and the diffusion of these samples go independently of the samples of charged particle; in the instant when they meet any sample of charged particle $s^{\s,\d},\ \s\in\{ +,-\},\ \d\in\{ r,i\}$ they in turn emit the new sample of this particle of the type $s^{\s ',\d},\ \s '=-sign(s)sign(co)\s$. We assume that the free photon samples propagate much faster than the samples of tied photons. 
The propagation of the scalar photon density is thus determined only by the operator of kinetic energy $E_{kin}=-\frac{\Delta}{2m}$ and we obtain that with the constant scalar photon emission rate $\frac{1}{2m}$ in the point $x_0$ in the short time frame their density will be equal to Green function $\rho (x)=-\frac{1}{4\pi\ 2m|x-x_0|}$ of the operator $E_{kin}$. Hence, after the transformation of scalar photon samples to the samples of probe particle located in the Coulomb field, we obtain the swarm behavior simulating one quantum particle in Coulomb field. The difference between free and tied photons is that the samples of free photons do not disappear, but create (emit) the samples of particles. 

The case of constant center $x_0$ is idealized because it was assumed that the particle created the field has the infinite mass and thus does not reveal quantum properties. A real particle (for example, proton in the hydrogen atom) must be considered as a quantum particle as well, and, correspondingly must be represented by its swarm of samples $d_j^{\s ,\d}$. Each of these samples must emit scalar photon samples as above. A system of two charged particles is then represented by three swarms:
\begin{equation}
\{ s_j^{\s ,\d},\ d_j^{\s ,\d},\  \bar\g , \bar g,\} ,\ \ j=1,2,\ldots;
\end{equation}
Standard quantum description of states in the form $\Psi(r_1,r_2)$ means that we consider the situation when $\Psi(r_1,r_2)=\Psi^{+,r}(r_1,r_2)-
\Psi^{-,r}(r_1,r_2)+i(\Psi^{+,i}(r_1,r_2)-\Psi^{-,i}(r_1,r_2)$ и $\Psi^{\s ,\d}(r_1,r_2)$ denotes the density of such pairs of samples of the form $s^{\s ' ,\d '},\ d^{\s '',\d ''}$ (relatively to the total number of such pairs over all  $\s ,\d$), such that one sample has transformed from the scalar photon sample emitted by the other sample and $(\s ,\d )=(\s ' ,\d ')(\s '',\d '')$ (product of pairs is defined by the natural way). The ordinary diffusion method Monte Carlo consists in that such pairs arise with intensity proportional to $1/|r_1-r_2|$. It is equivalent to that in the swarm approach any scalar photon sample $g$ emitted by one particle samples in turn emit (or absorb, dependently on the charge signs) the sample of the second particle in the moment when $g$ meets any sample of the second particle.  The swarm representation of wave functions then corresponds to the computations along the Monte Carlo scheme in the framework of Hilbert formalism. 

We show how Coulomb interaction can be reduced to the swarm dynamics defined above. Let the swarm $S$ corresponding to the charged particle be somehow divided to two parts: $S'$ и $S''$, which we treat as the physically different quantum particles. Let at first these two swarms are separated by some spatial distance $d$. We consider the free photon samples which emit (create) the samples of particles and move themselves only by the diffusion mechanism. The density of these free photon samples will then be distributed by the law $1/r$, where $r$ is the distance passed by a photon sample from the point of emission, because $1/r$ is proportional to Green function of the diffusion operator $\Delta$. If the photon samples emit (or absorb - dependently on the charge sign) the samples of the other particle swarm when they meet its members, it is equivalent to the situation when they are initially distributed as $1/r$, e.g., it is equivalent to Coulomb field.  

Hence the simple rule comes for finding of the approximation of Coulomb interaction. Coulomb interaction of two quantum particles is equivalent to the point wise interaction between their samples and the scalar photon samples emitted by the other particle samples, e.g., to the interaction which switches on when the coordinates of two samples corresponding to photon and particle coincide. We represent the quantum flow of one particle wave package through the tied photons transformations, which as the Coulomb free photon samples move via diffusion. Our representations of Coulomb field and free flow of wave package are then analogous. 
The accuracy of this approximation is limited by the supposition that the carriers of Coulomb field – free scalar photons samples have much higher diffusion rate than the tied photon samples which determine the action of the diffusion operator with the imaginary coefficient: $i\Delta$. The assumption of the point wise interaction makes swarm approach resembled to the density functional method, but this assumption in swarm representation will be valid in the three dimensional case only because only in this case Green function of the diffusion operator has the Coulomb form. In the other words, swarm method in the described form is applicable only to the Coulomb field in the three dimensional space. The supposition about point wise interaction radically simplifies the practical realization of swarm approach for many particles as well as the representation of entangled states. Really, in any step of simulation the number of operations will be $kN$ where $k$ is the total number of particles, $N$ - the total number of samples for one of them (and of photon samples emitted by its samples). The quadratic speedup comparatively with the search of all pairs "particle - particle" will be if we store in memory not the single photons but its density in all points of configuration space division.

How to account the identity of particles in the swarm approach? The general idea is to treat the identical particles as the different areas of the same swarm of samples. These areas can be separated, say, by the sign of sample $\s$ so that for the positive sign we treat the sample as belonging to the swarm of the first particle, for the negative – to the swarm of the second one. The Pauli principle is then guaranteed immediately. The more accurate consideration can be fulfilled if we introduce the spin for each sample of particles and make the intensity of interaction dependent on spins. 

\subsection{Transformation of tied photons to free photons}

The swarm representation of ordinary quantum mechanics requires the introduction of tied photon samples which can be emitted by and can transform to the samples of charged particles. The description of QED effects requires the criterion for the transformation of tied photon samples to the free photon which can be detected. Such a criterion can be based on the visual representation of quantum dynamics only because it cannot be described inside QED itself. The following topological criterion can be used. We call the division of swarm the situation when the spatial area occupied by this swarm is split to the different components of integrity: $D_1, D_2,\ldots, D_k$. One from the parts of swarm  $S_1,S_2,\ldots, S_k$, occupying these areas, say, $S_1$ is then called the swarm of particle and all other parts form free photons of vector type (vector photons) about which we assume that they are emitted by this particle. The tied photons emitted by swarms $S_2,S_3,\ldots,S_k$ yet do not transform to the samples of charged particle but simply change their type correspondingly to the cyclic rule $re,+\ \ar\ im,+\ \ar\ re,-\ \ar\ im,-\ \ar\ re,+$ and move with the constant speed corresponding to the impulse $p$ of the charged particle so that the wave function found by the summing of photon samples amplitudes of four types approximately equals $exp(ipx)$. Here we can separate the scalar photons propagating along the speed $v$ of initial wave package and the vector photons propagated along the perpendicular to $v$. 

The transformation of free photon samples to the samples of charged particle goes along the rules dual to ones defined above. A free photon sample $ph$ emits the sample of charged particle $s$ when $ph$ collides with a sample of $s$. This general receipt can be done in more details so that it will correspond to QED rules for the founding of amplitudes of emission and absorption of the virtual photons. For example, the probability of such transformation can be made proportional to the dot product of the polarization vector of $ph$ and the impulse of the sample $s$ it meets. The amplitude of emission (creation of photon sample by a sample of charged particle) and absorption (transformation of photon sample to a sample of charged particle) must coincide and must be proportional to the value
\begin{equation}
e\ i\ (\bar \e,\bar p),
\end{equation}
where $\bar \e$ and $\bar p$ are vector of photon polarization and impulse of charged particle, $e$ is the charge of electron, $i$ is imaginary unit. This rule of QED can be naturally reformulated for the swarm representation. This rule of QED can be naturally reformulated for the swarm representation. A photon sample arisen in the emission from some sample of charged particle of the type $\tau$ must have the type $(\tau+1)sign(\bar \e,\bar p)$, where the intensity of this emission must be proportional to the module of this scalar product. This rule gives the swarm approximation of QED. The analogous procedure takes place for the particles with spin (see \cite{Fe}). 

If the samples of charged particle are supplied with spins, we can introduce the spirality of free photon samples emitted by the samples $s$, of considered particle, so that the spirality of a photon sample $ph$ uniquely determines the spin of sample $s$ which emitted this photon sample. And vice versa, the photon spirality determines the spin of the sample of particle emitted by this photon sample when it meets any member of swarm $s$. The influence of the photon sample spirality will be opposite to the influence of spin of the sample emitting this photon sample. The closer is spin and spirality the easier will be the transformation from photon sample to the particle sample will be. We thus can simulate the effects connected with the interaction between spins of charged particles. 

\subsection{Representation of entangled states of many particle systems}

It was shown that the swarm method includes the dynamics of not entangled states. However, it is possible to extrapolate the swarm representation to the entangled states of the form (\ref{shm}). Let we are given a system of $k$ quantum particles. Its swarm description has the form:
\begin{equation}
\bar S^1,\bar S^2,\ldots,\bar S^k,\ \bar\g,\ \bar g,
\label{many}
\end{equation}
where $S^j$ is the swarm corresponding to the particle $j=1,2,\ldots,k$; $\bar\g,\ \bar g$ are the common swarms of tied and free photon samples correspondingly. For each $j$ we introduce the order on the swarm $S^j=\{ s^j_1,s^j_2,\ldots,s^j_{n_j}\}$ and will consider some pairs of samples of the form $s^j_l,s^i_l$ as connected ($i\neq j,\ i,j=1,2,\ldots,k$). We agree that if some sample disappears in the evolution then a sample of the other particle connected with it disappears as well. We then can simulate the entangled states of the form (\ref{shm}). Let any state $\Psi_j$ be the coordinate position of the particle. The order on the swarms can be defined so that exactly those samples will be connected which correspond to the wave functions of the form $\Psi^l_j$ and $\Psi^m_j$. We can also simulate the dynamics of such states by the changing of order on swarms.  

We can yet more simplify the representation if we require that exactly those pairs $s^j_l,s^i_l$ of samples are connected whose space-time coordinates coincide. At last we can suppose that the entanglement is transmitted by the photon sample from the sample which emitted it to the sample of the other particle. 

Coulomb interaction in the many particle system consists of point wise interaction between the samples of photons and particles. This interaction goes when the samples are located in the same point of space-time and with the intensity proportional to the probability of their collisions. Electromagnetic interaction goes through vector photons according the described rules. 

\section{Complexity of the simulation of Coulomb interaction and the speed of its propagation}

The physical characteristic of field – its speed of propagation is tightly connected with the complexity of simulation of the dynamics induced by this field. Let us consider once more swarm method of simulation of the dynamics of $k$ particle system which representation has the form (\ref{many}). In each step of the simulation we search all points of division of the configuration space, and for each point fulfill
\begin{itemize}
\item Step of emission of the photon samples by the samples of particles located in this point, from the swarms $S_j$, 
\item Step of the diffusion of the photon samples, located in this point, from the swarms $\bar\g,\ \bar g$, and 
\item Step of emission of particle samples by the photon samples, located in this point.
\end{itemize}

This method gives the complexity of simulation in each step of simulation of the order $kN$, where $N$ is the number of all elements of configuration space. Let us compare this method with the direct method of simulation of the  type “each sample of $i$-th particle with each sample of $j$-th particle” where the potential of interaction is proportional to $1/r(i,j)$ in case of Coulomb field, and interaction propagates with the speed $c$. It requires of the order $k^2N^2$ elementary operations to each step of iteration. To simulate one step of the direct method we need at most of the order $c$ steps of swarm method. Hence, if $c$ is finite, the complexity of collective behavior method has the order square root of the direct method complexity. But if $c$ is so large that $dr/c<dt$, where $dr,\ dt$ are the minimal sizes of the length and time in our model, our swarm method will work unsatisfactory. Really, if the field propagation in our representation is instantaneous, then the accuracy of our approximation will depend on the total number $k$, of particles, that violates the requirement of the scalability. We ignore the influence of particles located too far, but their summed influence in instantaneous interactions can be valuable. 

The requirement of the linear growth of computational recourses with the growth of $k$ leaded to the limitation of the speed of field propagation in the representation of collective behavior. The finite value of $c$ immediately gives the effect of delay. During the time required for the propagation of free photon samples, transmitting Coulomb field, the samples of tied photons change a little the shape of the particle wave function. It allows to simulate the relativistic effects in the method of collective behavior. 

We note that the large speed of free photon diffusion comparatively to the tied does not require the larger instantaneous speed of their moving. For this it would suffice to assume that that the intensity of creation of the samples of particles for the free photons is much smaller than for tied, but the intensity of their own creations is, in contrary, much higher. This relation makes possible to assume that the free photon samples transform to the samples of particles as the tied photon samples. 

The reasonings from above are valid for every field which propagation is determined by the local interactions of the type of cellular automata, as the diffusion method for Coulomb field. If algorithmic approach is universal and can embrace all the existed interactions in the common simulation algorithm, then we can roughly estimate the computational complexity of QED of electrons and atomic nuclei in comparison with the other interactions (strong, weak, and gravitational). If we suppose that the fraction of quantum non-locality in the common volume of computations is small, then the time of simulation $\tau$ will approximately correspond to the physical time. The computational complexity of QED will be then proportional to the time needed for the processing of the free photon samples shift to the unit of length, relative to the time needed for the processing of the same shift of the samples of charged particles, e.g. the fraction of speeds of particles and photons: $v_{part}/v_{phot}$. The large value of the speed of light $c$ in comparison with the typical speeds of charged particles serve as the indicator of the large computational complexity of the phenomena of not electrodynamic nature, which occupy the greater part of the time of the universal simulating algorithm. 

\section{Correspondence between swarm method and QED}

Swarm representation for many body systems with connected samples serves as a reduction of Hilbert formalism that is required by the algorithmic approach. This representation makes possible to build the films of processes which formulation in QED is incorrect: for example, the problem of capture of flying electron by a proton. Incorrectness of this problem in QED follows from that here the state of photon field is unknown, e.g., there are no photons at all. 
Hence, this problem cannot be solved through the scattering matrix. At the same time the inverse process - photo effect has the standard description in QED. Factually, the reason of incorrectness of such problems in QED is the general, and it follows from the principal impossibility to glue quantum mechanics with the classical. The difficulty is not in that it is impossible to join these two techniques by the unit set of formulas – it would join easily in the framework of the common simulating algorithm. The difficulty is that there is no criterion which determines for a given system whct technique to apply for it: classical or quantum. It explains the importance of the wave function collapse problem. The creation of a scalable quantum computer would not solve this difficulty but only reduces its sharpness because a quantum computer makes possible to extrapolate the existing methods to the wider class of problem without their substantial modification. But the absence of clear progress in the technologies of quantum processors gives grounds to suppose that that this difficulty is solved in Nature by some mechanism which immediate actions we observe in experiments but the explanation of it does not require to address to extremely complex objects\footnote{for example, to the consciousness of an observer (compare with \cite{Me}). The algorithmic approach, developed in this work is the search of such mechanism.}.
It gives the new sense to the attempts to find the modifications of QED which would be free of this difficulty, particularly, the modification based on the collective behavior. 

This task is substantially harder than the reduction of ordinary quantum mechanics to the collective behavior. At first, here we deal not with the logically simple and reliable quantum mechanics of Shoedinger equation, but with the logically incomplete quantum electrodynamics where such things as divergence of rows take place. At second, in the formalism we speculate on the incorrect in QED problems must have the certain solution. Factually, the passage
"quantum mechanics of many bodies <------> classical mechanics" must become the correct and monosemantic, that augurs much radical logical break with Hilbert formalism than the passage to the collective behavior in ordinary quantum mechanics. Classical algorithms, which earlier served at most as the replacement of the calculations by formulas, will now come in the conceptual part of description, and it is unavoidable if we intend to reach our main goal. It worsen dramatically perspectives to find the correspondence with QED. If for the ordinary quantum mechanics such correspondence can be lead to the form of mathematical theorems\footnote{Feynman path integrals give the equivalent form of quantum mechanics which does not touch its conceptual basement. But in the exact proof of this fact some suppositions must be introduced that are not evident beforehand: for example, about the uniform distribution of linear parts of paths and the complete substantiation requires certain efforts. But in case of new formalism which does not completely agree with Hilbert these difficulties grow many times. For the ordinary quantum mechanics of one particle this is not felt at all.}, then for the many body case with the field even the search of formulations of such theorems represents problems. It is impossible to “substantiate” completely our model in the framework of QED, because the divergence with QED is laid in the starting point of swarm model. The reasonable form of checking of all versions of algorithmic formalism including the method of collective behavior would be the description of the concrete processes and the comparison with the standard when the both descriptions are possible. The algorithmic approach pretends in perspective to the including to the physics the scientific areas which traditionally belong to the other sciences (chemistry), and the algorithmic representation of the dynamics presumes the visualization. But the structure of simulating algorithms must be simple enough to allow the comparison with the standard quantum description when the last is possible. It touches the scattering problems that are computed with high accuracy including relativistic corrections, etc. The requirement of good approximation in these cases is necessary for the development of our approach \footnote{The method of collective behavior cannot compete successfully with the analytical methods in any cases, because the real total number of samples can be much more than can be stored in the memory of our computers. But we expect that this method can substantially widen the borders of processors which we can simulate somehow. However, sometimes the methods of swarm type can beat analytical methods in the precision. The good example is the diffusion Monte Carlo method for the ground states of two electron system (for example, atom of helium or molecule of hydrogen).}.

\section{Conclusion}

We have defined the algorithm for the simulation of many particle quantum dynamics accounting the simplest EPR type of entanglement. This algorithm in the elementary form includes the Coulomb interaction only, but it can be supplied with the exchange of vector photons leading to the effects of QED. The main advantage of this algorithm is its scalability when the total number of particles grows. The further development of this algorithm is connected with the visual representation of the many particle dynamics. The visualization is the principal tool of the algorithmic approach which makes possible to generalize the method of theoretical physics on the complex many particle systems studied in chemistry and biochemistry.

\end{document}